\documentclass[fleqn, usenatbib]{mnras}
\usepackage{newtxtext,newtxmath}

\usepackage[T1]{fontenc}

\DeclareRobustCommand{\VAN}[3]{#2}
\let\VANthebibliography\thebibliography
\def\thebibliography{\DeclareRobustCommand{\VAN}[3]{##3}\VANthebibliography}

\usepackage{graphicx}	
\usepackage{amsmath}	

\title[Magnetic fields of hot exoplanets]{Magnetic Field Evolution of Hot Exoplanets}

\author[K. Kilmetis et al.]{
K. Kilmetis,$^{1}$\thanks{E-mail: kilmetis@strw.leidenuniv.nl}
A.~A. Vidotto$^{1}$,
A. Allan,$^{1}$
D. Kubyshkina,$^{2,3}$
\\
$^{1}$ Leiden Observatory, Leiden University, PO Box 9513, 2300 RA Leiden, The Netherlands, \\
$^{2}$ Space Research Institute, Austrian Academy of Sciences, Schmiedlstrasse 6, A-8042 Graz, Austria, \\
$^{3}$ Space Research and Planetology, Physics Institute, University of Bern, Sidlerstrasse 5, CH-3012 Bern, Switzerland
}

\date{Accepted XXX. Received YYY; in original form ZZZ}

\pubyear{2024}

\begin{document}
\label{firstpage}
\pagerange{\pageref{firstpage}--\pageref{lastpage}}
\maketitle

\begin{abstract}
Numerical simulations have shown that the strength of planetary magnetic fields depends on the convective energy flux emerging from planetary interiors. Here we model the interior structure of gas giant planets using \texttt{MESA}, to determine the convective energy flux that can drive the generation of magnetic field. This flux is then incorporated in the Christensen et al. dynamo formalism to estimate the maximum dipolar magnetic field $B^\mathrm{(max)}_\mathrm{dip}$ of our simulated planets. 
First, we explore how the surface field of intensely irradiated hot Jupiters ($\sim 300 M_\oplus$) and hot Neptunes ($\sim 20 M_\oplus$) evolve as they age. Assuming an orbital separation of 0.1 au, for the hot Jupiters, we find that $B^\mathrm{(max)}_\mathrm{dip}$ evolves from 240 G at 500 Myr to 120 G at 5~Gyr. For hot Neptunes, the magnetic field evolves from 11 G at young ages and dies out at $\gtrsim$ 2 Gyr.
Furthermore, we also investigate the effects of atmospheric mass fraction, atmospheric evaporation, orbital separations $\alpha$ and additional planetary masses on the derived $B^\mathrm{(max)}_\mathrm{dip}$. We found that  $B^\mathrm{(max)}_\mathrm{dip}$ increases with $\alpha$ for very close-in planets and  plateaus out after that. Higher atmospheric mass fractions lead in general to stronger surface fields, because they allow for more extensive dynamo regions and stronger convection.
\end{abstract}


\section{Introduction}
Most of the solar system planets and some of their moons possess  intrinsic magnetic fields \citep{schubert}, which are thought to arise from dynamo mechanisms in the planetary interior \citep[e.g.][]{fearndynamo, chistensenaubert}. Indeed, in-situ observations of magnetic fields have proven crucial in understanding the internal structure of solar system planets \citep{Banegalbook}. 

For a planet to maintain a magnetic field, magnetic energy has to be generated at timescales quicker than  diffusive timescales. This is achieved through a dynamo process, which requires three conditions to operate \citep{lazio_rev}. First, the planet has to be formed with a pre-existing magnetic field. 
This seed field is thought to originate during planetary formation, from the magnetized planet-forming disks \citep{seedfield}. Second, large-scale motions of electrically charged material inside an electrical conductor need to occur, allowing the induced electromotive force to generate a magnetic field. Such large-scale motions occur in planets both due to thermal gradients that arise due to cooling, as well as compositional gradients that arise due to differentiation of denser material that aggregates closer to the core \citep{differentiation}. Third, the flow must be non-symmetric, and more specifically, have a non-zero helicity \citep{helicity_moffat}, so that flow-lines are entangled, much like the magnetic flux-tubes that lead to coronal mass ejections in stellar surfaces. Should the above conditions be met, a planetary magnetic field is generated and can extend outside the surface of the planet, forming an intrinsic magnetosphere. 

Subsequent interactions between the planetary magnetosphere with the solar wind can trigger magnetic reconnection events both in the night and day sides of the planet, releasing energetic electrons that produce electron-cyclotron maser (ECM) emission \citep{ecmoriginal, ecm2}. For solar system planets, this radiation is detected at radio wavelengths \citep{zarka98}, and the (cyclotron) frequency of the emission depends on the strength of the magnetic field. Numerous attempts have been made to detect such emission from exoplanets \citep[e.g.]{oldradioattempt, oldtaubootis, griessmeier}, with some tentative detections being recently reported \citep[e.g.][]{hintofdetection, tbootisclosecall}. Several reasons can explain why most of the attempts to detect this emission so far resulted in non-detections. For example, the beamed emission requires a geometric alignment between the Earth and the ECM source \citep{kavanagh23}. Additionally, the power of the radiation is controlled by the strength of the stellar wind -- should the wind be weak or the system too far away, it is more challenging to observe the emission \citep{radioaline}. Another important condition for detecting ECM emission from exoplanets with ground-based telescopes is that the frequency needs to exceed the ionospheric cutoff at about 10 MHz \citep{ionosphere}. The cyclotron frequency dictates where the radio emission peaks, which is in turn set by the strength of the planetary magnetic field. Therefore, concrete estimates of exoplanetary magnetic fields are required in order to guide observational campaigns. 

The goal of our work is to provide constraints on exoplanetary magnetic fields. Here, we base our model in the work by \cite{chistensenaubert}, who simulated the dynamo process using realistic parameters describing the hydro-magnetic flow. Their work resulted in a scaling relation, mapping the internal properties of a planet to its generated magnetic field. Their models have accurately predicted the magnitude of the magnetic field in systems where the dipole term is dominant, like the Earth, Jupiter and rapidly rotating stars. For rapidly rotating systems, they showed that the magnitude of the planetary magnetic field depends only on the properties of the convective flux. 

The relation published by \cite{christensen-nature} has been used in several works to predict the strength of exoplanetary magnetic fields \citep{reiners_browndwarf, oldterrestialpaperthatscoopedus, yadav, mcintyre, hori}. An important consideration proposed by \citet{yadav} is that convective motions in the interior of close-in exoplanets are affected by the strong irradiation the planet receives from the host star. Therefore, the efficiency of magnetic field generation of highly irradiated exoplanets could change.  

In their work, \cite{yadav} combined the scaling law of  \cite{reiners_browndwarf}, which is a simplified version of the original \citeauthor{christensen-nature}'s scaling law,  with a model for the planetary interior to estimate magnetic fields of known hot Jupiters. The authors found that the incident flux can play a significant role in the generation of the magnetic field of the most massive hot Jupiters. Specifically, they calculated field strengths over 100 G, which are more than one order of magnitude higher than field strengths of the planets in the solar system. This is because the strong irradiation results in higher equilibrium temperatures, which change the temperature gradient that controls convection. \cite{yadav} did not find a clear relation between incident flux and magnetic field, though. This is because they had a diverse sample of exoplanets, making it more difficult to disentangle specific effects, such as due to mass, radius, orbital distances, on the predicted field strengths.

In this paper, we perform a parametric study to understand how  exoplanetary magnetism evolves, using a similar methodology as that proposed by \cite{yadav}. We consider planets of different masses (Jupiter-class versus Neptune-class), a range of initial envelope mass fraction and the distance from the host star. For lower mass planets, for which atmospheric escape can be significant, we also investigate the effects that atmospheric evaporation has on the dynamo process and by extension to the surface magnetic field. Additionally, for each combination of mass, atmospheric mass fraction and distance from the star, we compute the evolution of planetary magnetic field from 500 Myr to 10 Gyr. 

This paper is structured as follows. In section 2 we detail our methodology for simulating planetary interior sand their evolution and calculating the magnetic field. In section 3, we present the results of our numerical experiments. In section 4, we examine the assumptions that led to said results and place them in a wider context. Lastly, we summarize this work and draw conclusions in Section 5.

\section{Methods}
\label{sec:methods}
\subsection{Planetary Dynamos}
The dynamo process that generates planetary magnetic fields takes place within a dynamo region that extends from $R_\mathrm{dyn}^\mathrm{start}$ to $R_\mathrm{dyn}^\mathrm{end}$ in the interior of a planet. At the top of this region, the maximum magnetic field is predicted through the analytic relations presented in \cite{christensen-nature}:
\begin{equation}
    \label{eq:bdyn}
    B_\text{dyn}^\text{(max)} = c f_\mathrm{ohm} \langle \rho \rangle^{1/3} \left( F q_0 \right)^{2/3},  
\end{equation}
where $c$ is a proportionality constant set to 0.68, $f_\mathrm{ohm}$ is the ratio of ohmic dissipation and is set to 1, $\langle \rho \rangle$ is the average density of the dynamo region, and $q_0$ is the reference convective flux:
\begin{equation}
    q_0 = q_c(R_\text{dyn}^\mathrm{start}) \left( \frac{R_\text{dyn}^\mathrm{start}}{R_\text{dyn}^\mathrm{end}} \right)^2,
\end{equation}
where $q_c (R_\text{dyn}^\mathrm{start})$ is the convective flux computed at the start of the dynamo region. In equation (\ref{eq:bdyn}), $F$ is an efficiency factor that accounts for all radially varying features of the dynamo region, and is calculated as
\begin{equation}
    \label{eq:F}
    F^{2/3} = \int^{R_\mathrm{dyn}^\mathrm{end}}_{R_\mathrm{dyn}^\mathrm{start}} \left( \frac{q_\mathrm{c}(r)}{q_\mathrm{0}} \frac{l_\mathrm{conv}(r)}{H_\mathrm{T}}\right)^{2/3} \left( \frac{\rho(r)}{\langle \rho \rangle} \right)^{1/3} 4 \pi r^2 dr
\end{equation}
where $r$ is the radial coordinate, $q_\mathrm{c}(r)$ is the convective flux,  $l_\mathrm{conv}(r)$ is the convective length scale, $\rho(r)$ is the mass density and $H_\mathrm{T}(r)$ is the temperature length scale given by
\begin{equation}
\label{eq:ht}
    H_\mathrm{T}(r) = \frac{P(r)}{\rho(r) g(r) \nabla_\mathrm{adv}(r)}\, ,
\end{equation}
with $P(r)$ being the pressure, $g(r)$ the gravitational acceleration, and $\nabla_\mathrm{adv}$ the adiabatic, logarithmic gradient of temperature over pressure. In \cite{christensen-nature},  $F$ is assumed  to be of order unity for Earth, Jupiter, and main-sequence stars. For hot Jupiters and hot Neptunes, we found that $F$ can significantly deviate from this, with $F$ ranging from 10$^{-1}$ to 10$^{4}$, depending on the planet and its interior properties. The key quantity that impacts this is the convective flux $q_\mathrm{c}$. We calculate it as:
\begin{equation}
    \label{eq:qc}
    q_\mathrm{c}(r) = \frac{2c_\mathrm{p}(r)T(r)\rho^2(r) v_\mathrm{conv}^3(r)}{-P(r)\delta(r)},
\end{equation}
with $c_\mathrm{p}$ being the specific heat capacity at constant pressure, $v_\mathrm{conv}$ the velocity of convective motions, and $\delta$  is the derivative of $\ln\rho$ with respect to $\ln T$
at constant gas pressure.

To identify the regions where the dynamo process is active, we employ the criterion proposed by \cite{zhangandrogers} that uses the magnetic Reynolds number $\mathrm{Re}_\mathrm{mag}$. $\mathrm{Re}_\mathrm{mag} $ is an non-dimensional quantity that measures the effects of convection against magnetic diffusion. If it is greater than unity, convecting material generates a magnetic field in timescales shorter than the field diffusive timescale. Thus we do not only require the existence of convection but also it needs to be vigorous. Following \citet{zhangandrogers}, we consider areas to be dynamo-generating if 
\begin{equation}
    \label{eq:reynolds}
    \mathrm{Re}_\mathrm{mag} = \frac{\mathrm{convection}}{\mathrm{mag. diffusion}} = \frac{l_\mathrm{conv} v_\mathrm{conv}}{10^{-7}  T^{-3/2}\mu_0 }> 50,
\end{equation}
where $l_\mathrm{conv}$ is the characteristic length of a convective motion, $\mu_0$ is the magnetic permeability of the vacuum, and T is the local temperature, all given in cgs units. The threshold of $50$ is arbitrary, but we found that its exact value is irrelevant as once vigorous convection takes place, it discontinuously jumps from values much lower than unity to the tens of thousands. We define $R_\mathrm{dyn}^\mathrm{start}$ as the innermost radial shell of the planet where dynamo-generation is active and $R_\mathrm{dyn}^\mathrm{end}$ is the outermost radial shell. In equation (\ref{eq:reynolds}), we assumed that the magnetic diffusivity of plasma is $\mu_0/ \sigma$, where $\sigma [\textrm{s}^{-1}] \simeq 10^{7}  (T[\textrm{K}])^{3/2}$ is the electrical conductivity of the plasma \citep{tsinganos}. 

\begin{figure}
    \centering
    \includegraphics[width=\linewidth]{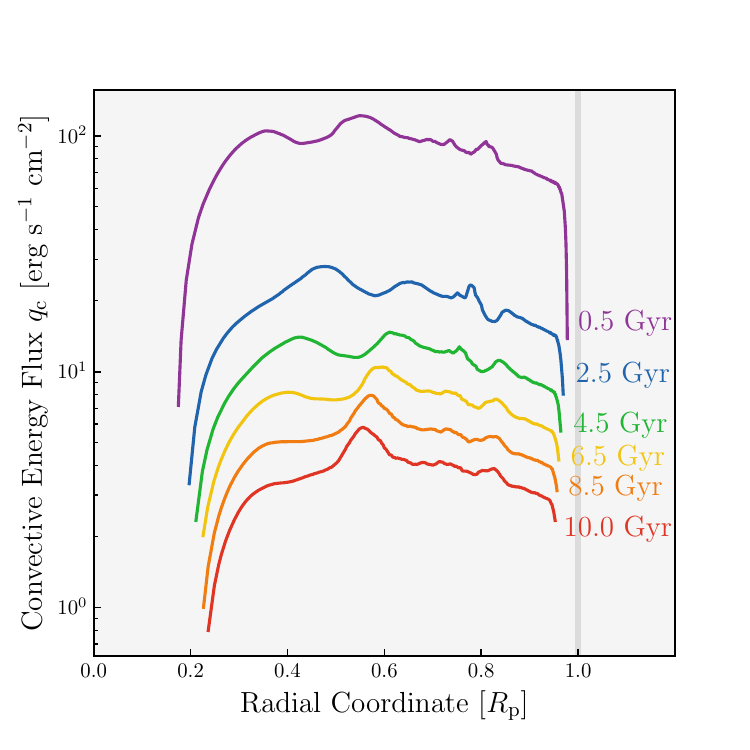}
    \caption{The convective energy flux in the active dynamo region, for a hot Jupiter with $M_\mathrm{p}$ = 317 $M_\oplus$, $\mathrm{f}_\mathrm{env}=$ 94\%, $\alpha=$ 0.1 AU, $s=$8 $k_\mathrm{B}$/baryon. Atmospheric evaporation is ignored. Different ages are shown with different colors.}
    \label{fig:qc}
\end{figure}

We examine the behaviour of the convective energy flux $q_c$ over the interior of a typical planet in figure \ref{fig:qc}, where we see that as the planet evolves, the behaviour of the convective energy flux is similar. It initially sharply increases up to a threshold that is more or less flat for the majority of the dynamo-active region. As the dynamo active region comes to its end, the convective energy flux quickly falls. As the planet evolves, the amount of total energy convected decreases. This decay is stronger early on the life of the planet -- as the planet sheds its initial heat from planetary formation we observe an order of magnitude decay in $q_c$. At later stages of its evolution, the convective flux decreases by factors of order unity.

To compute the magnetic field on the surface of planet $B_\mathrm{dip}^\mathrm{(max)}$, we assume that $B_\mathrm{dyn}^\mathrm{(max)}$ decays as a dipolar field, thus the magnitude of the field drops off with the inverse cube of the distance away from the dynamo surface \citep{reiners_browndwarf}. Thus
\begin{equation}
    \label{eq:bdip}
    B_\mathrm{dip}^\mathrm{(max)} = \frac{B_\mathrm{dyn}^\mathrm{(max)}}{\sqrt{2}} \left( \frac{R_\mathrm{p} - R_\mathrm{dyn}^\mathrm{end}}{R_\mathrm{p}} \right)^{3},
\end{equation}
with $R_\mathrm{p}$ being the radius of the planet. An important assumption of our model is that, by utilizing equation (\ref{eq:bdyn}), we implicitly assume that all planets we model are fast rotators. In this sense, all the values we calculate for the magnitude of the magnetic field are upper-limits and denoted with the `(max)' superscript. 

\subsection{Modeling Planets with \texttt{MESA}}
To simulate the evolution of planets, we utilize the code \texttt{Modules for Experiments in Stellar Astrophysics}, hereafter referred to as \texttt{MESA} (\citeauthor{mesa1}, \citeyear{mesa1}, \citeyear{mesa2}\footnote{details most of the software's capabilities regarding planets}, \citeyear{mesa3}, \citeyear{mesa4}, \citeyear{mesa5}  and \citealt{mesa6}). As the name suggests \texttt{MESA} specializes in stellar interiors, but it has also been used by the community to study planets to explore a wide range of topics \citep[e.g.,][]{chenrogersMESAoriginal, mesamalsky, dariaentropy}. For example,  \texttt{MESA} has been used to study Roche-lobe overflows in hot Jupiters \citep{mesavalsechii}, and to model the radius valley in hot sub Neptunes \citep{mesaowen}, proving the viability of using MESA in both planet regimes. \texttt{MESA} has also been used to model dynamical phenomena, such as the deposition of planetesimals on sub-Neptunes \citep{chatterjee}. 
Here, we follow the procedure detailed in  \cite{dariainsets}, which was adapted from the work of \cite{chenrogersMESAoriginal}. 
Throughout the development and testing of the code we employ here, its outputs  were compared to the results of some alternative models, reaching similar results to those of \cite{lopez14}, if assuming high initial temperatures of planets. Similarly, they were also consistent to results from the \texttt{JADE} (Joining Atmosphere and Dynamics for Exoplanets; \cite{attia21}) code and \texttt{Completo} (the planet evolution part of the Bern planet-formation code; e.g., \citealt{Emsenhuber21}), provided we use the same stellar input (see, e.g. \citealt{pezzottia, pezzottib}) and the planet does not experience significant migration. 
We consider planets through the lens of an one-dimensional, two-part model for their interiors. Each planet is constituted by an inert, rocky core and gaseous envelope with solar abundances. We solve the equations of structure only for the envelope using the core as an inner boundary condition. The outer boundary condition is dictated by the properties of the planet's host star and their orbital separation. 

Our planetary models are initialized by the following four parameters,
\begin{enumerate}
    \item Planetary mass  ($M_\mathrm{p}$)
    \item Initial atmospheric mass fraction ($f_\mathrm{env}$)
    \item Orbital separation ($\alpha$)
    \item Initial core specific entropy ($s$)
\end{enumerate}
For the vast majority of our similations we vary the first three parameters, electing to assign the same value for the inital core entropy for every planet, this value being $s$ = 8 $k_\mathrm{b}$/baryon. While changes in the energy flux emerging from the planet are of interest to us because of their important role in the generation of magnetic field, changes due to different initial entropy values \citep{dariaentropy} are mostly limited to the first hundred Myr of its existence. After the energy in the interior of the planet redistributes itself and the system ``forgets'' its initial conditions, alterations on the initial entropy of a planet impact the magnetic field in a small manner. For these reasons, we chose to track the evolution of planets past the first 500 Myr. We discuss the impact of entropy on the final magnetic field in section \ref{sec:discussion}.

Aside from the characteristics of the planetary interior, our model allows us to define the properties of the host star and the planet's orbit. We assume that all planets are in circular orbits around a solar-type star. The star's mass is kept at 1 $M_\odot$ and its initial rotation period - which defines the intensity of high energy radiation - is set at 7 days.  Following \cite{dariaentropy}, the evolution of the star (e.g., its luminosity, radius and effective temperature) is tracked with the \texttt{Mors} package \citep{mors}. As a benchmark, the effective temperature of our planet-hosting star at 5.5 Gyr is 5849 K.
Additionally, \texttt{Mors} can also calculate the star's X-ray and ultraviolet (XUV) fluxes estimated from the parameters derived in the rotational evolution model by \cite{stellarmodels}. Here, we allow the orbital separation to range from 0.01 au to 2 au. These are the orbital separations for which most exoplanets have been discovered, and additionally for distances beyond this range the differences in the equilibrium temperature do not impact magnetic field generation. 

\texttt{MESA} can provide a variety of outputs in the form of \texttt{history} and \texttt{profile} files. The former gives a macroscopic overview of the planet at a given time while the latter provides radial profiles of all the quantities that define the planet, at a given time. Because of our need to integrate over a specific region to calculate the efficiency factor of equation \ref{eq:F}, we exclusively make use of \texttt{profile} files. In table \ref{tab:symbols} we provide a map between default \texttt{MESA profile} outputs and the mathematical symbols we use here.

\begin{table}
\centering
\caption{Mapping between default \texttt{MESA} profile outputs and our adopted mathematical notation. The first four default outputs are in logarithmic form, so we use $x = 10^{\log x}$ to derive their linear values (second column). \label{tab:symbols} }
\begin{tabular}{ll}
\hline
\texttt{MESA}                      & Symbol                         \\ \hline
\texttt{logR}                      & $r$                            \\
\texttt{logRho}                    & $\rho(r)$                      \\
\texttt{logT}                      & $T(r)$                         \\
\texttt{logP}                      & $P(r)$                         \\
\texttt{grav}                      & $g(r)$                         \\
\texttt{mlt\_mixing\_length}       & $l_\mathrm{conv}$              \\
\texttt{conv\_vel}                 & $v_\mathrm{conv}$              \\
\texttt{dlnRho\_dlnT\_const\_Pgas} & $\delta$                       \\
\texttt{grada}                     & $\nabla_\mathrm{adv}$          \\
\texttt{cp}                        & $c_\mathrm{p}$ \\
\hline
\end{tabular}
\end{table}
We split the planets we model into two categories, Jupiter-class and Neptune-class, which occupy two very distinct regions of the $M_\mathrm{p}-f_\mathrm{env}$ parameter space. As Neptune-class, we consider planets whose mass can range from 7 M$_\oplus$ to 27 M$_\oplus$ alongside an initial atmospheric mass fraction of 2\% to 10\%. Jupiter-class planets possess masses in the range 150 M$_\oplus$ to 500 M$_\oplus$  and their $f_\mathrm{env}$ can be 86\% to 94\%. The mass of the inert core is then set by the planetary mass and the fraction of which is in the atmosphere, $M_\mathrm{core} = M_\mathrm{p}\left(1-f_\mathrm{env}\right)$. The core radius is determined by its uniform density, which we set to 4 g/cm$^3$ following \cite{DENSITYrogers}.

\section{Results: Computation of exoplanetary magnetic fields}
\subsection{Magnetic field evolution with planetary age}
\begin{figure}
    \centering
    \includegraphics[width = \columnwidth]{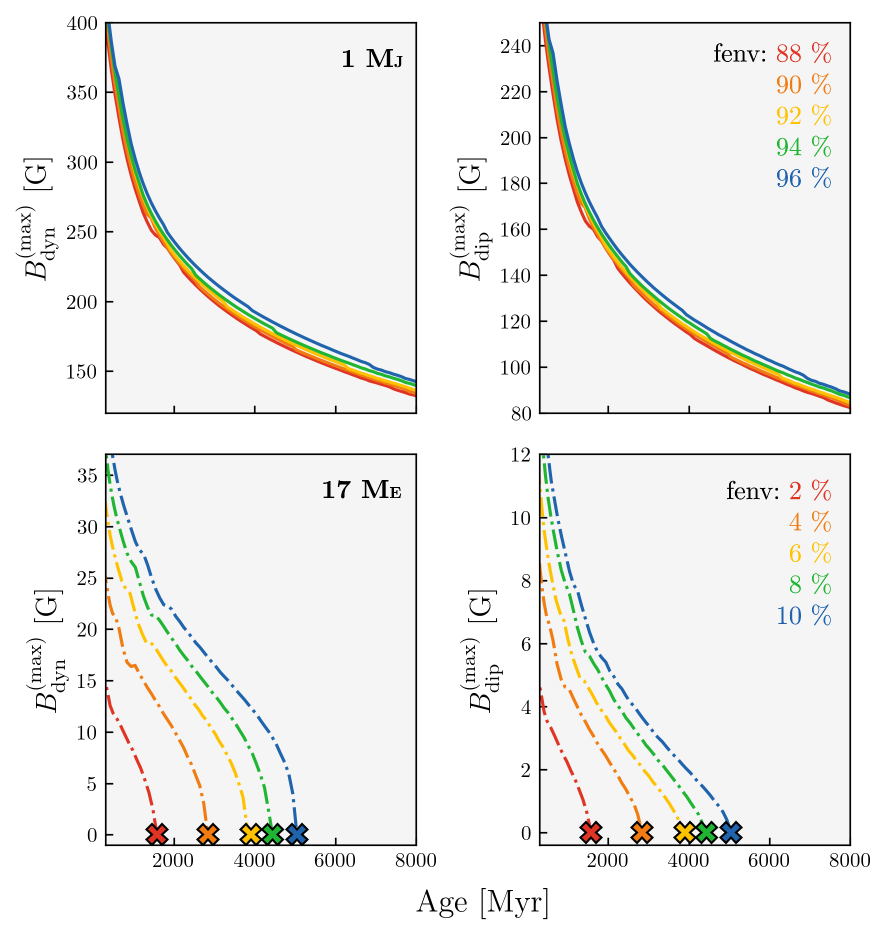}
    \caption{Top row: Evolving magnetic fields of a typical hot Jupiter, of mass 1 M$_\mathrm{J}$ and varying initial atmospheric envelope fractions.  The left panels showcases the temporal evolution of the magnitude of the magnetic field in the planetary surface ($B_\mathrm{dip}^\mathrm{(max)}$). The right panels showcase the temporal evolution of the magnitude of the magnetic field in the dynamo surface ($B_\mathrm{dyn}^\mathrm{(max)}$).
    Bottom row: Evolving magnetic fields of a typical hot Neptune, of mass 17 M$_\oplus$ and varying initial atmospheric envelope fractions. Notice, the y-axis scaling varies in all panels. The 'X' markers show when the dynamo ceases to operate.
    }
    \label{fig:typical}
\end{figure}

First, we investigate the evolution of the magnetic field. Figure \ref{fig:typical} showcases the magnetic fields of a prototypical hot Jupiter (top panels) and a prototypical hot Neptune (bottom panels). The hot Jupiter has a total mass of 1 M$_\mathrm{J}$ (317 M$_\oplus$), the hot Neptune has a total mass of 17 M$_\oplus$. We initialize them with varying initial atmospheric fractions, and they orbit at 0.1 au. For now, we do not account for the effects of atmospheric evaporation, which will be investigated in Section \ref{subsec:evap}.

Figure \ref{fig:typical} shows a monotonic decrease of the dynamo (left) and surface (right) fields with time. This decay occurs because as planets age, they cool down and their luminosities, which trace their convective flux, wanes. Because in our model magnetic fields are coupled to the properties of the convective flux, the predicted magnetic field strengths decay as the planet ages.

As we increase the amount of mass initially contained in the atmosphere, our planets generate stronger magnetic fields. For a given age, the field strengths increase a few G for  higher $f_\mathrm{env}$, for  both Jupiter-class and Neptune-class planets. Higher atmospheric envelope fractions imply that more material is available for convection, which yields stronger fields due to a more extended dynamo region. This region is larger in both relative (larger fraction of the planet is dynamo active) and absolute terms with increase of $f_\mathrm{env}$. Additionally the vigor of convection, on average, tends to increase with $f_\mathrm{env}$.

For the Neptune-class, the dynamo dies after a few Gyr. The time for which a dynamo remains operational on such planets depends on the initial envelope fraction. The dynamo stops working because, as the planet cools, convection becomes less efficient in comparison to magnetic diffusion, thus our magnetic Reynolds number criterion $\left( \mathrm{Re}_\mathrm{mag}>50 \right)$ is not met anywhere in the planetary interior. The dynamo regions of the Jupiter-class planets also exhibit a decay as they evolve, but their dynamo regions start off larger and hence, the magnetic Reynolds number criterion is always met throughout the their evolution.

\subsection{Magnetic field dependence with orbital separation and mass}
\begin{figure}
    \centering
    \includegraphics[width = \columnwidth]{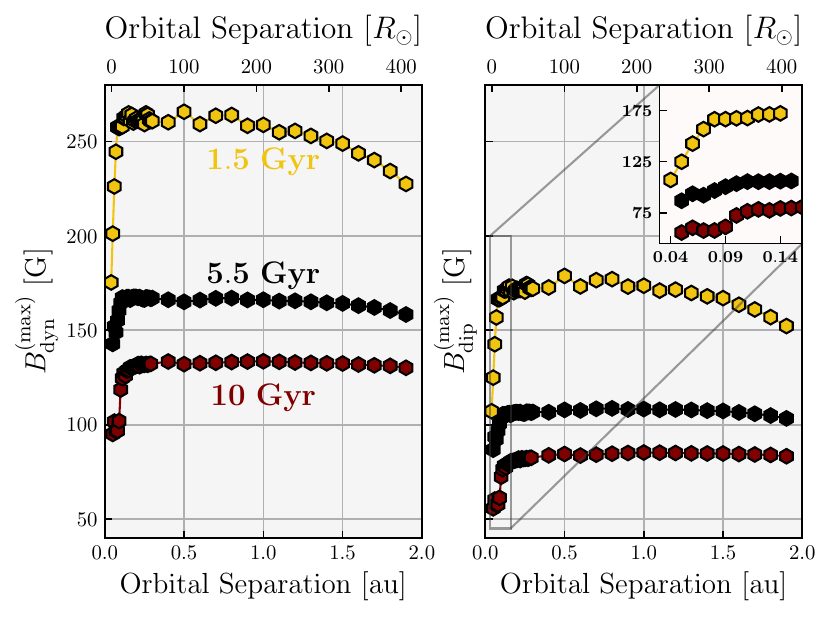}
    \caption{Dynamo (left) and surface dipole (right) magnetic fields of a Jupiter-class planet of 317 M$_\oplus$, f$_\mathrm{env}$ = 94$\%$, for varying orbital separations at three ages, marked with colors. The inset shows the surface dipole, for small orbital separations.}
    \label{fig:jup-orbsep}
\end{figure}

We examine how the impact of orbital separation on the surface magnetic field changes as a planet evolves. For now, we focus our attention on a single planetary model, that of a Jupiter-class with a mass of 317 $M_\oplus$, f$_\mathrm{env}$ = 94\%. We present the evolution of strength of the dynamo (left) and the dipole (right) fields as a function of orbital separation in Figure \ref{fig:jup-orbsep}.

For the plotted ages, both  $B_\mathrm{dyn}^\mathrm{(max)}$ (left panel) and $B_\mathrm{dip}^\mathrm{(max)}$ (right panel) are subdued for very short orbital separations. Within this close orbital separations (see inset in the Figure) the increase in field strengths is steeper for younger planets versus older ones. The fact that the field strength is reduced for extremely irradiated planets is because they are puffier and thus have a lower average density, as per equation (\ref{eq:bdyn}).

As we increase the orbital distances, we see that the field peaks and this peak moves further away for older planets -- from 0.08 au for 1.5 Gyr-old planets to 0.13 au for the 10-Gyr old ones. This effect correlates with the increasing bolometric luminosity stars output as they evolve. 

For 1.5 Gyr-old hot Jupiters, we observe that the surface magnetic field decreases past the threshold value as orbital separation further increases. This decrease is less evident for Hot Jupiters aged 5.5 Gyr and almost disappears for 10 Gyr old planets. This occurs because planets further away from their host star can shed the heat of planetary formation quicker than their closer-in counterparts. After the dissipation of this initial energy occurs, the magnetic field is less dependent on the orbital separation for $\alpha \gtrsim 0.2$ au.

\begin{figure}
    \centering
    \includegraphics[width = \columnwidth]{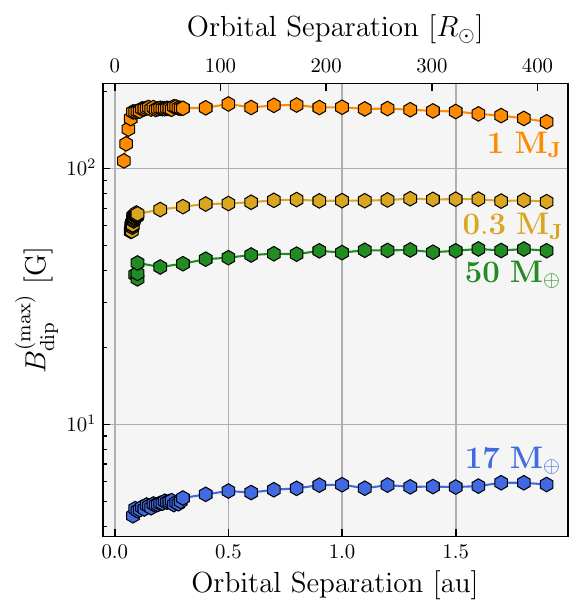}
    \caption{Surface magnetic fields of planets of differing mass at 1.5 Gyr for varying orbital separations. Orange and blue are typical Jupiter-class and Neptune-class planets with f$_\mathrm{env}$ = 94$\%$ and f$_\mathrm{env}$ = 6$\%$. Yellow is a Saturn mass planet with f$_\mathrm{env}$ = 90$\%$. Green is a super-Neptune with f$_\mathrm{env}$ = 70$\%$. The planetary masses are annotated in the figure.}
    \label{fig:orbsep}
\end{figure}

In Figure \ref{fig:orbsep}, we extend our investigation to planets with different masses and envelope mass fraction. Figure \ref{fig:orbsep} presents the archetypal Jupiter-class planet alongside with a 17-$M_\oplus$ mass planet, with an initial atmospheric fraction of 6$\%$ (corresponding to a core mass of 16 $M_\oplus$). In addition, we also simulate a Saturn-like planet with 0.3 M$_\mathrm{J}$ (95 M$_\oplus$), f$_\mathrm{env}$ = 90$\%$ and a super-Neptune of 50 M$_\oplus$, f$_\mathrm{env}$ = 70$\%$. In figure \ref{fig:orbsep}, we present the magnetic field for an age of 1.5 Gyr. The surface field increases monotonically but not linearly with planetary mass. The surface magnetic field exhibits an increase until 0.5 au for all planetary classes. Past 0.5 au and until 2 au, we observe qualitative differences for different planetary masses. The hot Jupiter's field decays, the hot Saturn's stays approximately constant, and the fields of both the hot super-Neptune and the hot Neptune exhibit marginal increases. Quantitatively, the values of the field does not diverge substantially from the one at 0.5 au.

\subsection{Expanding our parametric study}
\begin{figure*}
    \centering
    \includegraphics[width = 0.7\textwidth]{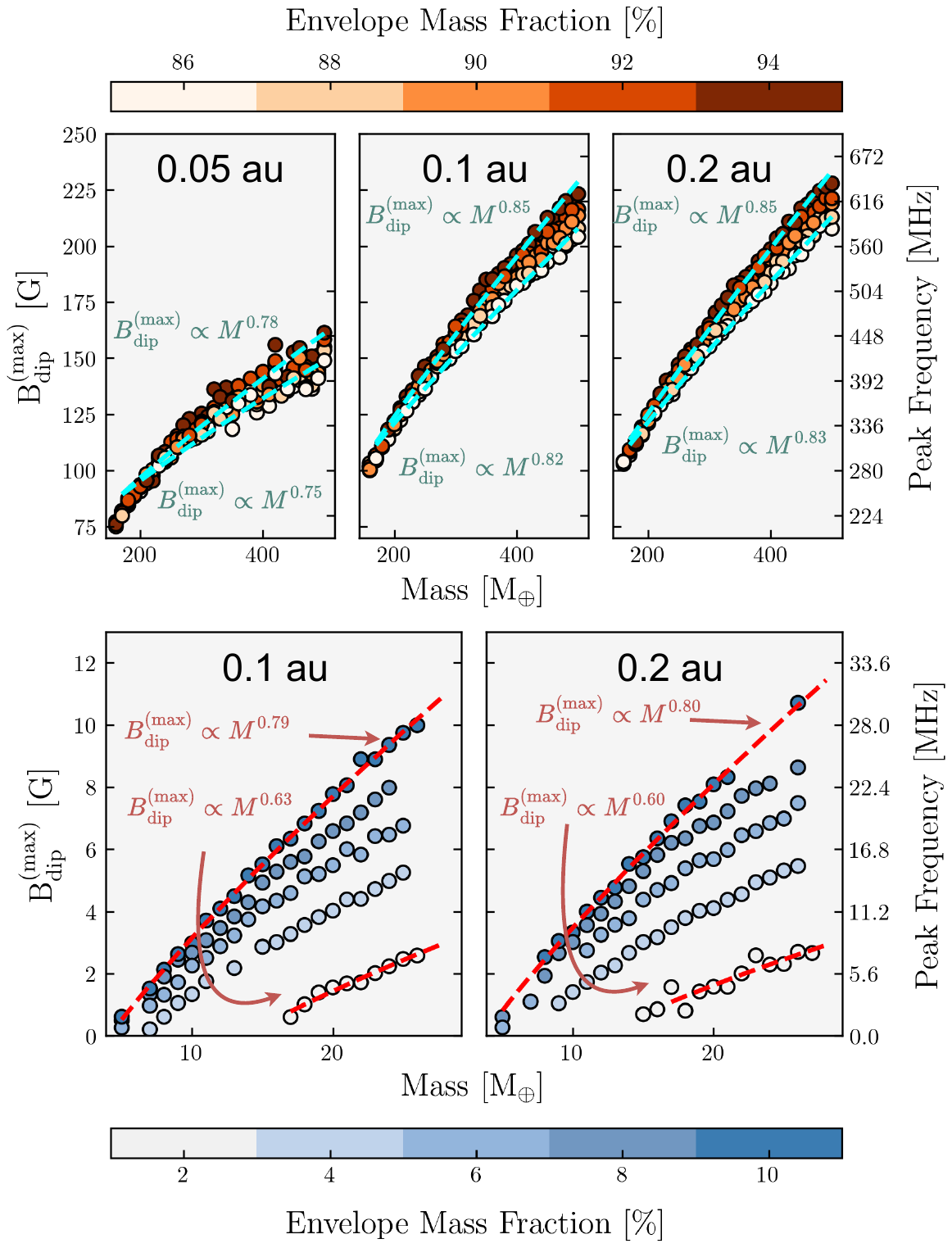}
    \caption{Maximum surface magnetic field strength vs planetary mass, at 1.5 Gyr. The right y-axis measures the peak emmision frequency as per equation \ref{eq:peak}. The upper panels depict Jupiter-class planets, while the lower panels depict Neptune-class planets. The orbital separation increases from the left to right. The colors correspond to atmospheric envelope mass fraction. The Jupiter-class planets are mapped in orange, the Neptune-class planets' envelope fraction are mapped in  blue.
    The dashed lines are power-law fits for the lowest and highest envelope fraction in each panels.
    }
    \label{fig:big-one}
\end{figure*}

The strength of our methodology is its speed. Exploiting that, we simulated $\sim$2000 of planets to investigate the effects of varying mass, envelope mass fraction, and orbital separation. In Table \ref{tab:sample} we present the inputs and results of selected models.The Jupiter-class planets we simulated range in mass from 150 M$_\oplus$ to 500 M$_\oplus$. Most of their mass is in their gaseous envelope, made with solar abundances. Their atmospheric envelope mass fraction ranges from 86\% to 94\%, which we explore in increments of 2\%. Additionally, we explore Neptune-class planets, whose atmospheric envelope mass fraction we vary from 2\% to 10\%. Their masses  range from 7 M$_\oplus$ to 27 M$_\oplus$. The calculated surface magnetic fields is presented in figure \ref{fig:big-one} for Jupiter class (upper row) and Neptune class (lower row) planets. We observe qualitative similarities between the two planet classes. In both cases their magnetic fields exhibit increasing behaviour with both planetary mass and atmospheric envelope fraction.

\begin{table} 
\centering
\caption{Selected numerical experiments. The first three columns are inputs to our models (planetary mass, envelope mass fraction, orbital separation, respectively), while the last three columns are outputs of our runs (age, the field strength at the top of the dynamo region, and the dipolar surface field strength, respectively). In all simulations below we did not include atmospheric evaporation and used the same initial specific entropy of 8 $k_\mathrm{b}/\mathrm{baryon}$.}
 \label{tab:sample}
\begin{tabular}{llllll}
\hline
$M_\mathrm{p}$ $[M_\oplus]$ & ${f}_{\rm env}$ [\%] & \multicolumn{1}{l|}{$\alpha$ {[}au{]}} & age {[}Gyr{]} & $B_\mathrm{dyn}$ {[}G{]}    & $B_\mathrm{dip}$ {[}G{]} \\ \hline
17  & 2  & \multicolumn{1}{l|}{0.1}  & 1 & 7   & 2   \\
17  & 2  & \multicolumn{1}{l|}{0.1}  & 4 & 0      & 0      \\
17  & 2  & \multicolumn{1}{l|}{0.1}  & 7 & 0      & 0      \\
17  & 2  & \multicolumn{1}{l|}{0.2}  & 1 & 7      & 2   \\
17  & 2  & \multicolumn{1}{l|}{0.2}  & 4 & 0      & 0      \\
17  & 2  & \multicolumn{1}{l|}{0.2}  & 7 & 0      & 0      \\
17  & 10 & \multicolumn{1}{l|}{0.1}  & 1 & 29  & 9   \\
17  & 10 & \multicolumn{1}{l|}{0.1}  & 4 & 12  & 2   \\
17  & 10 & \multicolumn{1}{l|}{0.1}  & 7 & 0      & 0      \\
17  & 10 & \multicolumn{1}{l|}{0.2}  & 1 & 29  & 9   \\
17  & 10 & \multicolumn{1}{l|}{0.2}  & 4 & 12  & 2   \\
17  & 10 & \multicolumn{1}{l|}{0.2}  & 7 & 0      & 0      \\
150 & 86 & \multicolumn{1}{l|}{0.05} & 1 & 146 & 78  \\
150 & 86 & \multicolumn{1}{l|}{0.05} & 4 & 103 & 53  \\
150 & 86 & \multicolumn{1}{l|}{0.05} & 7 & 80  & 40  \\
150 & 86 & \multicolumn{1}{l|}{0.2}  & 1 & 188 & 115 \\
150 & 86 & \multicolumn{1}{l|}{0.2}  & 4 & 116 & 68  \\
150 & 86 & \multicolumn{1}{l|}{0.2}  & 7 & 93  & 52  \\
150 & 94 & \multicolumn{1}{l|}{0.05} & 1 & 143 & 75  \\
150 & 94 & \multicolumn{1}{l|}{0.05} & 4 & 103 & 52  \\
150 & 94 & \multicolumn{1}{l|}{0.2}  & 7 & 82  & 40  \\
150 & 94 & \multicolumn{1}{l|}{0.2}  & 1 & 183 & 111 \\
150 & 94 & \multicolumn{1}{l|}{0.2}  & 4 & 120 & 69  \\
150 & 94 & \multicolumn{1}{l|}{0.2}  & 7 & 96  & 53  \\
400 & 86 & \multicolumn{1}{l|}{0.05} & 1 & 227 & 146 \\
400 & 86 & \multicolumn{1}{l|}{0.05} & 4 & 168 & 107 \\
400 & 86 & \multicolumn{1}{l|}{0.05} & 7 & 133 & 83  \\
400 & 86 & \multicolumn{1}{l|}{0.2}  & 1 & 316 & 212 \\
400 & 86 & \multicolumn{1}{l|}{0.2}  & 4 & 197 & 129 \\
400 & 86 & \multicolumn{1}{l|}{0.2}  & 7 & 160 & 103 \\
400 & 94 & \multicolumn{1}{l|}{0.05} & 1 & 237 & 152 \\
400 & 94 & \multicolumn{1}{l|}{0.05} & 4 & 180 & 114 \\
400 & 94 & 0.05                      & 7 & 143 & 90  \\
400 & 94 & \multicolumn{1}{l|}{0.2}  & 1 & 343 & 230 \\
400 & 94 & \multicolumn{1}{l|}{0.2}  & 4 & 216 & 141 \\
400 & 94 & \multicolumn{1}{l|}{0.2}  & 7 & 176 & 114 \\
\hline
\end{tabular}
\end{table}

For both classes of planets, 
a greater $f_\mathrm{env}$ leads to more material being available for convection and thus larger dynamo-active regions. Increasing $f_\mathrm{env}$ has more dramatic effect for higher mass planets, of any class, as a greater amount of mass is added to the hydrogen/helium envelope.

To quantify how the fields of planets with equal $f_\mathrm{env}$ change as their mass increases, we fitted power-laws of the form $B_\text{dip}^\text{(max)} \propto M_\mathrm{p}^{A}$, to the planets of the highest and the lowest $f_\mathrm{env}$ -- that is, for 86\% and 94\% for Jupiter-class planets, and 2\% and 10\% for the Neptune-class. These power law fits are shown in figure \ref{fig:big-one}. While the exponents increase with orbital separation (fits become steeper), they have similar trends for different $f_\mathrm{env}$ for Jupiter-class planets. On the other hand, for Neptune-class planets, $B_\mathrm{dip}^\mathrm{(max)}$ increases more steeply for higher $f_\mathrm{env}$. This is because for Neptune-class planets the volume the gaseous atmosphere occupies varies greatly for different $f_\mathrm{env}$. After it dominates the planet, volume-wise, the way the magnetic field scales with planetary mass stays the same. The most gaseous Neptune-class planets behave similarly to the least gaseous Jupiter-class planets.

In principle, magnetic fields such as the ones we calculate, could be observable in the radio wavelengths via auroral emission \citep{zarka01}. We can predict the frequency in which such cyclotron emission would peak as:
\begin{equation}
    \label{eq:peak}
    f_\mathrm{peak} = 2.8 \hspace{2pt} \frac{B_\mathrm{dip}^\mathrm{(max)}}{1 \mathrm{G}} \hspace{2pt} \mathrm{MHz}.
\end{equation}
The peak frequency is shown on the second y-axis of figure \ref{fig:big-one}. All of the Jupiter-class planets have magnetic fields large enough to generate radiation whose peak frequency exceeds the Earth's ionospheric cutoff of 10 MHz. The same holds for the Neptune-class planets with $M$ > 15 $M_\oplus$ and $f_\mathrm{env} $> 4\%.  

This does not mean that the planets could generate auroral emission observable by ground based facilities. To achieve this, the planets need to generate a sufficiently strong signal. The power of the signal depends on the size of the magnetosphere, which scales with the magnitude of the surface field, but also with factors that do not pertain to the planet, such as the distance between the Earth and the system, the density of the stellar wind, and the relative velocity between the stellar wind \citep{radioaline}. 

\subsection{Influence of atmospheric evaporation on predicted field strengths}
\label{subsec:evap}
Lastly, we investigate how the surface magnetic field is impacted by different atmospheric evaporation prescriptions. We are able to calculate the amount of mass lost due to extreme UV and X-ray photoevaporation. As the planets lose material, they shrink. In our numerical setup, we calculate this loss through both a realistic hydrodynamic approximation and the energy-limited approximation \citep{EL1, EL2}. A detailed account of the implementation of the hydrodynamic evaporation we use can be found in \cite{hydroevap}.

\begin{figure}
    \centering
    \includegraphics[width = \columnwidth]{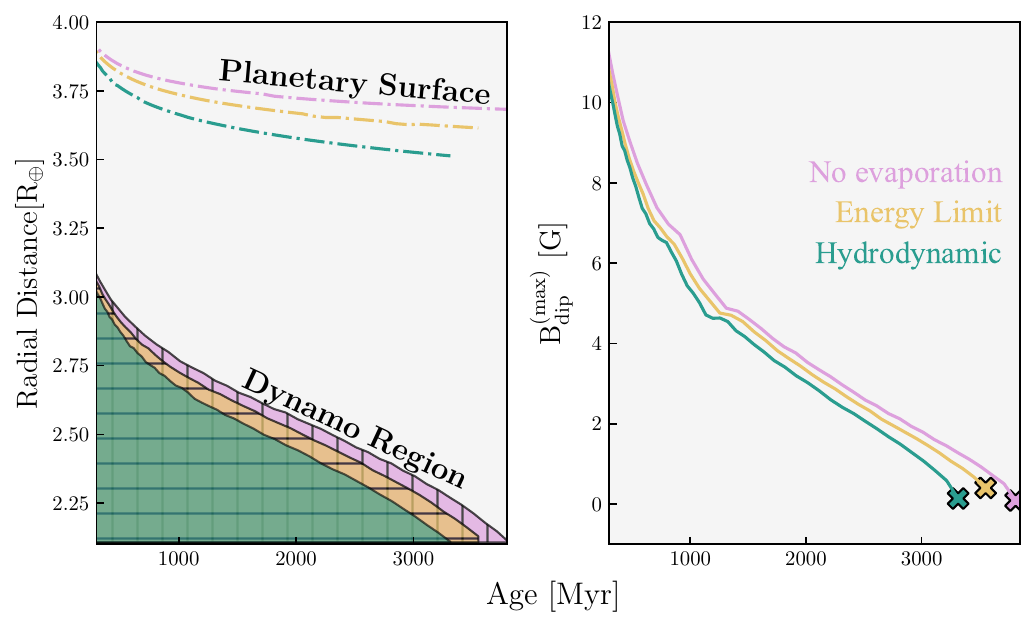}
    \caption{Left panel: evolution of the planetary surface (lines) and dynamo active regions (colored regions). Right panel: surface magnetic fields. Here, we consider a Neptune-class planet of 17 M$_\oplus$, f$_\mathrm{env}$ = 6$\%$, $\alpha $ = 0.1 au, under different atmospheric evaporation regimes: pink \& vertical hatching: no evaporation; yellow \& horizontal hatching: energy limited approximation; teal: hydrodynamic evaporation.}
    \label{fig:evaporation}
\end{figure}

In figure \ref{fig:evaporation}, we present the temporal evolution of the dynamo region and the dipolar surface field as a function of time. We examine the evolution of a Neptune-class planet of 17 M$_\oplus$, f$_\mathrm{env}$ = 6$\%$ , $\alpha$ = 0.1 au. We do not examine Jupiter-class planets as the total mass lost due to evaporation is small compared to their total mass. As a result, evaporation on Jupiter-class planets has negligible effects on the strength of the magnetic field. We  show the evolution of the planet without evaporation (pink line), using the energy-limited approximation (yellow) and the hydrodynamic approximation (teal).

In all cases, the dynamo-active region begins by encompassing about 25\% of the planet and shrinks over time, similarly to the shrinkage in planetary radius. 
Both the size of the dynamo-active region and the total size of the planet shrink faster as we move from the least sophisticated treatment of atmospheric evaporation (i.e., no evaporation) to the most sophisticated one (hydrodynamic aproximation), as the mass-loss rate increases for the more realistic approximations. This increased rate leads to a quicker dynamo death earlier (i.e., in these cases, $\mathrm{Re}_\mathrm{mag} <$ 50 throughout the planetary interior), while the smaller dynamo regions, inevitably lead to weaker surface magnetic fields. The magnetic fields of the three planets which we show here begin as potentially observable from ground-based radio observatories but quickly drop below the Earth's ionospheric cutoff, for $\lesssim$ 5 G. For this specific set of planets, only younger ($\lesssim 1$Gyr) Neptune-class planets could be observable through their auroral emission using ground-base observations.

\section{Discussion and model limitations}
\label{sec:discussion}

\subsection{Defining the dynamo region}
\begin{figure}
    \centering
    \includegraphics[width = \columnwidth]{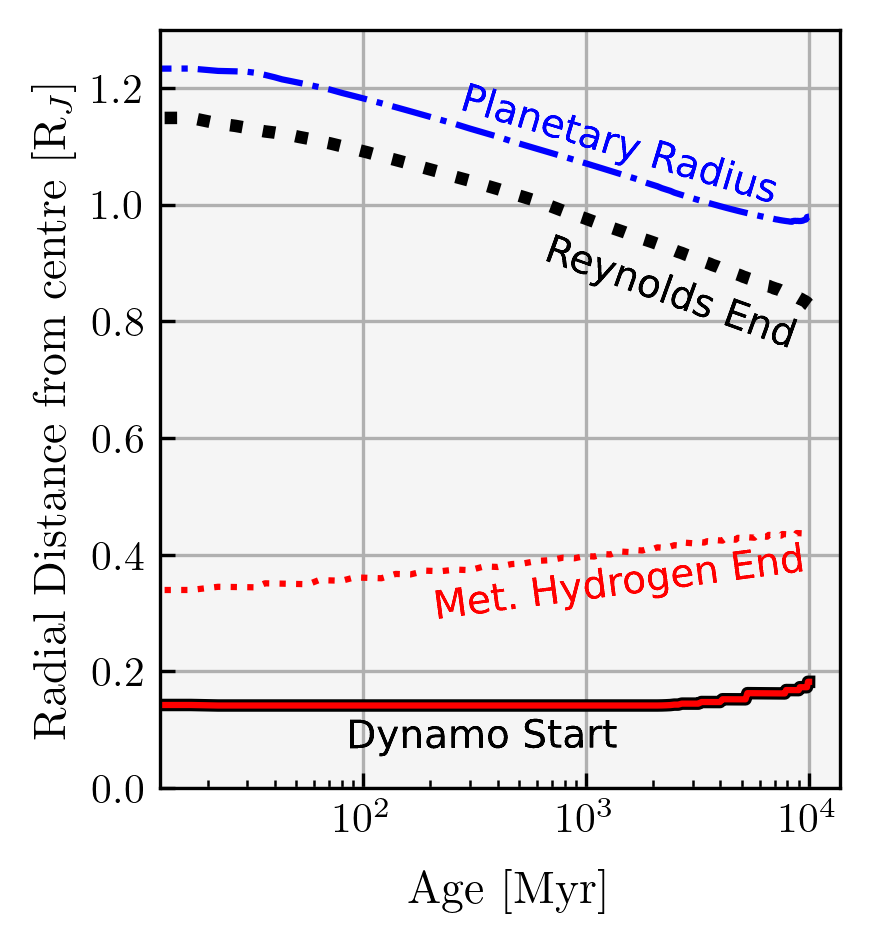}
    \caption{The size of the dynamo region for a planet with mass 95$M_\oplus$, initial envelope fraction of 95$\%$, 0.1 au away from a G-type star, through two differing criteria for the identification of the dynamo region: using the magnetic Reynolds number $<50$ (black) and metallic hydrogen (red). The solid lines mark the beginning of the dynamo regions and the dotted ones the end of them. The area below the solid line corresponds to the inert planetary core assumed in our MESA models. The planetary surface radius is the dot-dashed blue line. }
    \label{fig:dynregion}
\end{figure}

The magnitude of the fields we calculated here is unprecedented in the solar system planets. Jupiter's magnetic field averages at 5.5 G \citep{yujupB} and Neptunes at 0.37 G \citep{neptunefield}. Our  calculate surface fields for hot Jupiters are about 100 G, and of the same order of magnitude as those calculated by  \cite{yadav}. \cite{hori}, following a similar methodology as in our work, arrived at values around 50--100 G for a 1-M$_\mathrm{J}$ planet orbiting close to its host star. Our approaches differ on how our models set the boundaries of the dynamo-active region.
\cite{hori}, focusing exclusively on Jupiter-class planets, assumed that the dynamo region coincides with the region where metallic hydrogen exists. In their model, any regions of the planet that satisfies the necessary temperature and pressure conditions for hydrogen to exist in this phase are assumed to generate a dynamo. In our model, in contrast, we use a criterion based on the Reynolds number. 
Figure \ref{fig:dynregion}  compares these two criteria, where we see that the Reynolds criterion results in  larger dynamo regions (from ``Dynamo Start'' until ``Reynolds End'') when compared to the metallic hydrogen criterion (from ``Dynamo Start'' until ``Met.~Hydrogen End''). In our Reynolds criterion, the size of the dynamo region decreases as the planet evolves. In the metallic hydrogen criterion from \citet{hori}, the dynamo region increases with time, even though, it remains substantially smaller than that computed using the Reynolds criterion. 

We implemented the metallic hydrogen criterion used in \cite{hori} in our Jupiter-class planets and found that $B_\mathrm{dip}^\mathrm{(max)}$ is 94\% smaller than that derived with the $\mathrm{Re}_\mathrm{mag}$ criterion. Neptune-class planets never reach the pressures and temperatures necessary to host metallic hydrogen on their cores, so we did not perform this comparison.

\subsection{Effects of initial conditions (entropy)}
\begin{figure}
    \centering
    \includegraphics[width = 0.75\columnwidth]{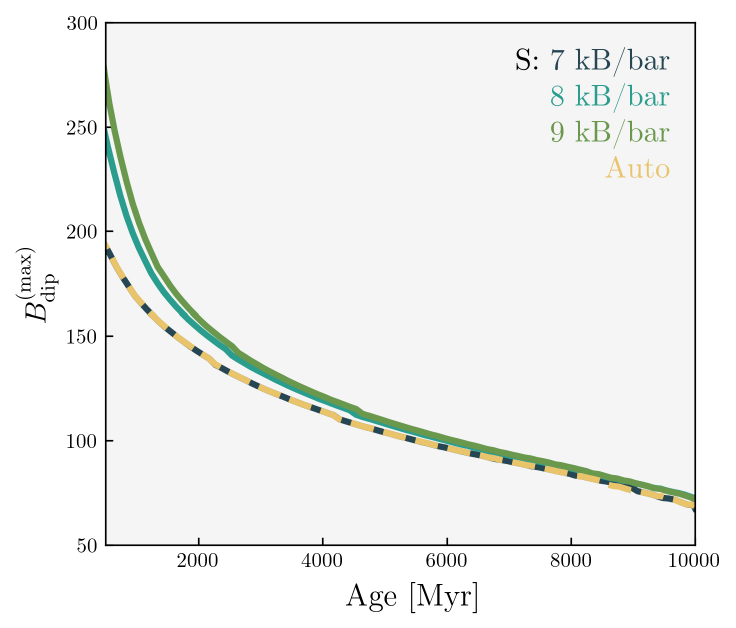}
    \caption{The evolution of the surface dipole field for different initial entropies. This is done for a 317 M$_\oplus$, $f_\mathrm{env}$ = 94\%, 0.1 au Jupiter-class planet.}
    \label{fig:entropy}
\end{figure}

In our calculations we assumed the same initial entropy for every planet, regardless of mass and orbital distance. We examine this assumption in figure \ref{fig:entropy}. We simulated an archetypal Jupiter-class planet 317 M$_\oplus$, $f_\mathrm{env}$ 94\%, $\alpha$ = 0.1 au with four different initial entropies for the gaseous envelope at 7,8,9 $k_\mathrm{B}$/baryon. We also have the option to let \texttt{MESA} designate an appropriate value for the initial entropy (``AUTO''). We find that changing the entropy does not significantly change the magnitude of the surface magnetic field, albeit it has a large impact on the efficiency of the convection in the planetary interior. This stems from the change in initial entropy impacting the convective flux but not impacting the size of the dynamo region.

The effects in the magnetic field are small, but they are non-linear. Increasing the entropy from 7 to 8 $k_\mathrm{B}$/baryon results in an increase in the magnetic field magnitude across the entire evolution, but further increasing it to 9 $k_\mathrm{B}$/baryon reduces the magnetic field. The ``AUTO'' value \texttt{MESA} chooses lies very close to 7 $k_\mathrm{B}$/baryon. Regardless, the effect of different initial entropies in our magnetic field estimates is minor, and they become less important with evolution. This is also the case for Neptune-class planets.

\subsection{Other caveats and speculative scenarios}
We note that our models apply only to planets composed out of an extensive, solar composition envelope and a metallic core. Our models cannot, for example, model planets with compositions similar to the solar system's ice giants with metal-rich mantles \citep{fornteyandnettelman}. Because Neptune's and Uranus' magnetic fields are generated in metallic mantles instead of a convecting hydrogen/helium envelope, it is not dominated by a dipolar field \citep{uranepsims}. The \cite{chistensenaubert} relations, therefore, do not apply to such a system. 

Aside from more metallic envelopes, differences in the composition of the dynamo-active region can lead to differences in magnetic field generation. For example, using \texttt{MESA} \cite{guchen} showed that the deuterium fraction of evaporating sub-Neptunes increases significantly over time. Since deuterium has a different density and heat capacity than the composition assumed here, the average density and heat capacity of the envelope would change, yielding a different convective flux and thus a different magnetic field. Moreover, any density gradients in the envelope, like the helium rain in Jupiter \citep{yamillajup}, could augment  magnetic field generation.

We elected to keep the same host star, a G-class dwarf, for all of our simulations. However, our model, through \texttt{Mors}, can be used to simulate planets around stars of spectral type F, K, and M by changing the mass of the host star. A change in mass induces a change in bolometric luminosity, which provides a different boundary value for convection. 
Similar effects can be mimicked by changing 
the orbital separation of the planets. In our experiments, we observed that strong stellar fluxes tend to reduce the dynamo strength. With that in mind, we speculate that
the same planet with the same orbital separation orbiting smaller stars with weaker fluxes could in principle generate stronger magnetic fields.
Likewise, we speculate that an inwards migrating planet would generate a weaker field, as it spends more time in orbits closer to its host star (i.e., receiving stronger stellar fluxes). 

Further, the code \texttt{Mors} allows for the consideration of more active stars, which are modeled assuming a faster rotation rate. Faster rotating stars have stronger XUV fluxes, which can lead to more evaporation and thus quicker ``dynamo deaths'', as we explored in Figure \ref{fig:evaporation}. Quantifying the sensitivity of the dynamo to a changing orbital separation while also alternating the properties of the host star is a topic meriting further study. Both of these could, in principle, be tackled with \texttt{MESA}.

Finally, we highlight again that our models inherit the assumptions from \citeauthor{christensen-nature}'s model, in which it is assumed that the energy ﬂux determines the magnetic ﬁeld strength in rapidly rotating objects. For this reason, the values we calculated here for the magnitude of the magnetic field should be considered as upper limits, similar to findings derived for fast stellar rotators \citep{2022A&A...662A..41R}.

\section{Conclusions}
We conducted one-dimensional planetary structure simulations by utilizing a version of \texttt{MESA} capable of coupling the evolution of the planet with that of the host star's \citep{chenrogersMESAoriginal, dariainsets}. For all our planets we utilized a two-part interior model composed of an inert core plus a gaseous envelope. We calculated the maximum magnetic surface field through scaling relations \citep{christensen-nature} that model the generation of a magnetic field through a dynamo process driven by the interior convective flux. 

We identify the region of the planetary interior within which the dynamo operates, by enforcing $\mathrm{Re}_\mathrm{mag}$>50. We explored the available parameter space by varying the planetary mass, atmospheric mass fraction, orbital separation and atmospheric evaporation prescription. From these, we conclude the following:
\begin{itemize}
    \item The predicted magnetic fields decay as planets evolve. Hot Neptunes and hot Jupiters generate  stronger maximum fields than those in the  solar system planets. For hot Jupiters, we find field strengths varying from $\sim 240$ G at 500 Myr to $\sim 90$ G at 8 Gyr. For hot Neptunes, typical field strengths evolve from 11 G to 0 G, from 500 Myr to 4 Gyr.  See Figures \ref{fig:typical} and \ref{fig:jup-orbsep}.
    \item A greater initial atmospheric mass fraction monotonically leads to greater surface magnetic fields. This effect is most prominent as the total mass of the planet increases. See Figure \ref{fig:big-one}.
    \item The surface magnetic field weakens for planets extremely close to their host star. For planets with greater orbital separations, the surface magnetic increases up to a threshold. Afterwards, it stays on that threshold until it approximately stabilises for distances beyond 1 au. This effect diminishes as the planet ages. See Figures \ref{fig:jup-orbsep} and \ref{fig:orbsep}.
    \item Atmospheric evaporation does not affect magnetic field generation in hot Jupiters. For hot Neptunes, however, realistic treatment of atmospheric evaporation leads to greater mass loss and consequently less material available for convection. These planets then consistently produce weaker fields and their dynamo ``dies'' in shorter timescales. By ``dynamo death'' we refer to the dynamo region being characterized by $\mathrm{Re}_\mathrm{mag}$<50. 
\end{itemize}
Our treatment is able to predict the magnetic field of a vast array of planets. We focus on Jupiter-class  and Neptune-class planets. The area between super-Neptunes and sub-Jupiters still needs to be explored. 

In our work, we found that Jupiter-class planets can produce maximum magnetic fields whose possible auroral emission would peak in frequencies above the Earth's ionospheric cutoff. This means that they could be detectable with ground-based telescopes. For Neptune-class planets, their magnetic fields could only be detectable with ground-based observatories at earlier ages ($\lesssim 600$~Myr).

\section*{Acknowledgments}
APA and AAV acknowledge funding from the European Research Council (ERC) under the European Union’s Horizon 2020 research and innovation programme (grant agreement No 817540, ASTROFLOW). We thank U.R. Christensen and A. Reiners for answering our questions regarding the scaling relations they derived. We thank other colleagues from our group Dr.~F. Driessen, Dr.~D. Evensberget, and Dr.~S. Belotti for discussions over this project.
We thank the anonymous referee for their thorough analysis, insightful comments and suggestions that have greatly enhanced this work.

\section*{Data Availability}
\texttt{MESA} inlists used are available at \url{https://zenodo.org/records/4022393} \citep{kubyshkina_daria_2020_4022393}. The post-processing code is available at \url{https://github.com/KKilmetis8/ExoMag}. The data described in this article will be shared on reasonable request to the corresponding author.
\bibliographystyle{mnras}

\bsp	
\label{lastpage}
\end{document}